\documentstyle[aps,epsbox,preprint]{revtex}

\title{Density matrix renormalization group for
       the Berezinskii-Kosterlitz-Thouless transition of the
       19-vertex model}
\author{Yasushi Honda and Tsuyoshi Horiguchi}
\address{Department of Computer and Mathematical Sciences,
         Graduate School of Information Sciences, \\
         Tohoku University, Sendai 980-77, Japan}

\newcounter{posix}
\newcounter{posiy}
\newcommand{\harrow}[3]{
   \setcounter{posix}{#1}
   \setcounter{posiy}{#2}
   \put(\value{posix},\value{posiy}){\line(1,0){4}}

   \addtocounter{posix}{6}
   \put(\value{posix},\value{posiy}){\circle{4}}
   \put(\value{posix},\value{posiy}){\makebox(0,0){#3}}

   \addtocounter{posix}{2}
   \put(\value{posix},\value{posiy}){\line(1,0){4}}
}

\newcommand{\varrow}[3]{
   \setcounter{posix}{#1}
   \setcounter{posiy}{#2}
   \put(\value{posix},\value{posiy}){\line(0,1){4}}

   \addtocounter{posiy}{6}
   \put(\value{posix},\value{posiy}){\circle{4}}
   \put(\value{posix},\value{posiy}){\makebox(0,0){#3}}

   \addtocounter{posiy}{2}
   \put(\value{posix},\value{posiy}){\line(0,1){4}}
}

\newcommand{\varrowb}[3]{
   \setcounter{posix}{#1}
   \setcounter{posiy}{#2}
   \put(\value{posix},\value{posiy}){\line(0,1){4}}

   \addtocounter{posix}{-2}
   \addtocounter{posiy}{4}
   \put(\value{posix},\value{posiy}){\framebox(4,4){#3}}

   \addtocounter{posix}{2}
   \addtocounter{posiy}{4}
   \put(\value{posix},\value{posiy}){\line(0,1){4}}
}

\newcommand{\harrowb}[3]{
   \setcounter{posix}{#1}
   \setcounter{posiy}{#2}
   \put(\value{posix},\value{posiy}){\line(1,0){4}}

   \addtocounter{posiy}{-2}
   \addtocounter{posix}{4}
   \put(\value{posix},\value{posiy}){\framebox(4,4){#3}}

   \addtocounter{posix}{4}
   \addtocounter{posiy}{2}
   \put(\value{posix},\value{posiy}){\line(1,0){4}}
}

\begin{document}
\tightenlines
\maketitle

\begin{abstract}
  We embody the density matrix renormalization group (DMRG) method
for the 19-vertex model on a square lattice
in order to investigate the Berezinskii-Kosterlitz-Thouless
transition.
Elements of the transfer matrix of the 19-vertex model are
classified in terms of the total value of arrows in one layer of
the square lattice.
By using this classification, we succeed to reduce enormously
the dimension of the
matrix which has to be diagonalized in the DMRG method .
We apply our method to the 19-vertex model with the interaction
$K=1.0866$ and
obtain
 $c=1.006(1)$ for the conformal anomaly.
\
\end{abstract}
\narrowtext

\section{Introduction}
The density matrix renormalization group (DMRG) method is developed
to obtain eigenvalues of Hamiltonian matrix for one-dimensional
quantum systems \cite{White}.
This method enables us to investigate a finite system with a large
size by a numerical calculation of matrices,
which can be handled within
recent computer resources such as a memory size
and a cpu speed.
Recently the DMRG method has been applied to the transfer matrix of
classical spin systems \cite{Nishino}.
In the both cases of quantum systems and classical systems,
each spin variable has a discrete degree of freedom.
Therefore we have matrices with a finite dimension
for these finite systems.
On the other hand, the dimension of the transfer matrix
becomes infinite for a finite system of spins
with continuous degree of freedom
such as a classical XY model.
Fortunately it is known that the XY model on a square lattice
$\Lambda$ is translated into a 19-vertex model for which the transfer
matrix is described in terms of realization of arrow variables
\cite{Knops}.
An arrow variable takes three discrete states.
Hence the 19-vertex model is called a three-state vertex model.
We can construct the transfer matrix with a finite dimension for
a finite system of the 19-vertex model
described by the three-state variables.
The 19-vertex model is solved by Zamolodchikov and Fateev
\cite{Zamo} in the case that its Boltzmann weights satisfy the
Yang-Baxter relation.
This exact solution is generalized to the $q$-state vertex
model by Sogo et al. \cite{Sogo1,Sogo2}.
When the Yang-Baxter relation is not satisfied,
we do not have an exact solution yet for that system.
However it is believed that a critical behavior of
the 19-vertex model without
frustration belongs to the same universality
class as the Berezinskii-Kosterlitz-Thouless transition
\cite{Knops,Bere,Kos1,Kos2}.

The purpose of the present paper is to embody the DMRG method for the
19-vertex model and to show that the dimension of matrices,
which are diagonalized in the DMRG method, is reduced enormously
by using the ice rule of the
19-vertex model \cite{Baxtertext}.
We obtain a value of the
conformal anomaly as $c=1.006(1)$ at $K=1.0866$ which is
regarded as the BKT transition point \cite{Knops}.
This value is consistent with a value of $c$
expected at the BKT transition point.
On the other hand, the value of $\eta/2$ appears to be $0.1175(5)$,
which is smaller than the expected value $\frac{1}{8}$
at the BKT transition point.
This suggests that the critical value of $K$ for the 19-vertex model
is smaller than 1.0866.

In section \ref{sec:19-vertex}, we briefly explain the relation
between the classical XY model and the 19-vertex model.
In section \ref{sec:dmrg}, we explain our method by which
we can reduce the dimension of transfer matrix in the DMRG method.
In section \ref{sec:results}, we show our results and give discussions
about the conformal anomaly and the smallest scaling
dimension.
Section \ref{sec:summary} is for a summary of the present study.

\section{ 19-vertex model }
\label{sec:19-vertex}
The partition function of the 19-vertex model is derived from that
of the XY model.
Assigning Boltzmann weights to 19 vertices, the 19-vertex
model describes the XY model on a square lattice
for each case with frustrations or without frustrations.
For spin variables, which take continuous values like classical
XY spins or plane rotators,
we have a transfer matrix with infinite dimension even for a finite
system.
Therefore the bare XY model cannot be handled by a numerical
diagonalization of the transfer matrix.
On the other hand, the 19-vertex model is described
by discrete variables which express
three kinds of arrows.
Hence we can make a transfer matrix with a finite dimension and
apply the DMRG procedure for a finite system.
In order to demonstrate the efficiency of our approach to the BKT
transition, we apply it to the 19-vertex
model without frustration for which the critical behavior of the
usual BKT transition has to be reproduced.

Let us briefly explain a relation between the XY model and the
19-vertex model on a square lattice $\Lambda$.
The partition function $Z_{\rm XY}$ of the XY model on the square
lattice $\Lambda$ is defined as follows:
\begin{equation}
Z_{\rm XY} = \prod_{k \in \Lambda} \int_{-\pi}^{+\pi} d \theta_k
  \exp\{ K \sum_{\langle i,j \rangle}
         \cos( \theta_i - \theta_j - A_{ij})
       \}  ,
\label{eq:xyptf}
\end{equation}
where $i,j$ and $k$ denote site indices, $\theta_i$ an angle of
the XY
spin at site $i$, $K$ an interaction parameter, respectively.
The sum $\sum_{\langle i,j \rangle}$ is taken over all
nearest-neighbor pairs of sites.
A bond parameter related to frustration
between sites $i$ and $j$ is expressed by $A_{ij}$.
The frustration $f$ is defined in terms of the $A_{ij}$ as follows:
\begin{equation}
  f \equiv \frac1{2\pi} \sum_{\rm P} A_{ij} ,
\end{equation}
where the summation $\sum_{\rm P}$ is taken over an elementary
plaquette.
If the value of $f$ is a half odd integer, the plaquette has a
frustration.

In a region of small parameter $K$,
we expand the exponential function in the
partition function for the XY model given in Eq.(\ref{eq:xyptf}).
Hence let us start with a following partition function $Z$:
\begin{eqnarray}
  Z &=& \prod_{k \in \Lambda} \int_{-\pi}^{+\pi} d \theta_k
        \prod_{\langle i,j \rangle}
        \{
           1 + K \cos(\theta_i - \theta_j - A_{ij})
        \} \nonumber \\
    &=& \prod_{k \in \Lambda} \int_{-\pi}^{+\pi} d \theta_k
        \prod_{\langle i,j \rangle}
        \left\{
           1 + \frac{K}{2} \exp \{ i(\theta_i-\theta_j-A_{ij}) \}
             + \frac{K}{2} \exp \{-i(\theta_i-\theta_j-A_{ij}) \}
        \right\} .
\label{eq:expptf}
\end{eqnarray}
The integrand of the  partition function $Z$ has the U(1) symmetry
as well as that of $Z_{\rm XY}$.
The first term of the integrand in rhs of
Eq.(\ref{eq:expptf}) is assigned to no arrow in the
19-vertex model.
The second term is assigned to an arrow pointing from a site $i$ to $j$
and the third term to
an arrow pointing from a site $j$ to $i$.
We use a word "arrow" even for the case of no arrow in a bond.
Only when the value of arrows incoming to a site is equal to the
value of arrows outgoing from the site, a weight of the arrow
configuration for the site survives after taking
integrations with respect to an angle
of XY spin at the site.
Otherwise the weights of arrow configurations do not appear in the
partition function.
In the six-vertex model, we have a similar rule to that mentioned
above \cite{Baxtertext}.
That rule is called as the ice rule, whose name comes from the
property of hydrogen ions in an ice crystal.

\begin{figure}
\begin{center}
\setlength{\unitlength}{0.01\textwidth}
\begin{picture}(24,24)

  \harrow{0}{12}{$\alpha_i$}
  \varrow{12}{12}{$\beta_i$}
  \harrow{12}{12}{$\gamma_i$}
  \varrow{12}{0}{$\delta_i$}

  \put(13,14){\makebox(0,0){$i$}}
\end{picture}
\end{center}
\caption{Four arrows surrounding a site $i$ \label{fig:fourv}}
\end{figure}
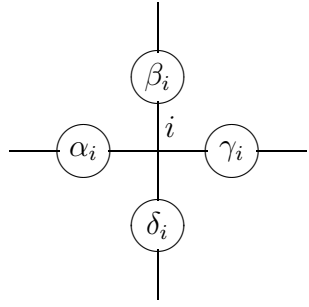

The 19 kinds of arrow configurations on vertices are permitted by an
"ice rule", which is a generalization of the ice rule for the
six-vertex model, when we include no arrow on a bond as shown
in Fig.\ref{fig:19vertex}.
Hereafter we simply say the 19 kinds of vertices instead of the 19
kinds of arrow configurations on vertices.
The vertex weight $W(v_i)$ depends on the kind of the vertex
$v_i \in \{1,2, \cdots, 19\}$ at site $i$.
The value of $v_i$ is determined by a configuration of four arrows as
follows:
\begin{equation}
  v_i = v_i(\alpha_i, \beta_i, \gamma_i, \delta_i) ,
  \label{eq:vertex}
\end{equation}
where $\alpha_i, \beta_i, \gamma_i$ and $\delta_i$ denote
arrows surrounding the site $i$ as shown in Fig.\ref{fig:fourv}.
Let us express an up and a right arrow by $+1$, a down and
a left arrow by
$-1$, no arrow by $0$, respectively.
For instance $v_i(-1,0,0,+1)=1$ as shown in Fig.\ref{fig:19vertex}.
The ice rule is described in terms of
$\alpha_i, \beta_i, \gamma_i$ and $\delta_i$
as follows:
\begin{equation}
  \alpha_i-\beta_i-\gamma_i+\delta_i = 0 .
\label{eq:ice}
\end{equation}
Using the weights for 19 vertices, we can describe the partition
function $Z$ as follows:
\begin{equation}
  Z={\sum_{\{ v_i \}}}' \prod_{i \in \Lambda} W(v_i) ,
\label{eq:19vertexptf}
\end{equation}
where the summation is taken over all permitted configurations of the
vertex on the lattice $\Lambda$.

\section{Density matrix renormalization group method with
         restrictions on the total value of arrows }
\label{sec:dmrg}
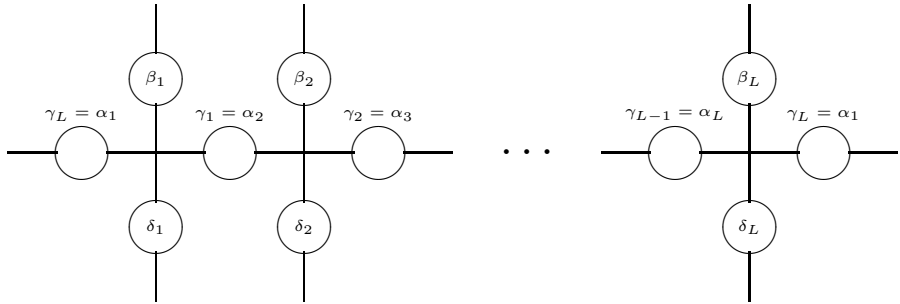
\begin{figure}
\begin{center}
\setlength{\unitlength}{0.01\textwidth}
\begin{picture}(72,24)
  \tiny
  \harrow{0}{12}{}
  \put(6,15){\makebox(0,0){$\gamma_L=\alpha_1$}}
  \varrow{12}{12}{$\beta_1$}
  \harrow{12}{12}{}
  \put(18,15){\makebox(0,0){$\gamma_1=\alpha_2$}}
  \varrow{12}{0}{$\delta_1$}

  \varrow{24}{12}{$\beta_2$}
  \harrow{24}{12}{}
  \put(30,15){\makebox(0,0){$\gamma_2=\alpha_3$}}
  \varrow{24}{0}{$\delta_2$}

  \put(42,12){\makebox(0,0){\Large $\cdots$}}

  \harrow{48}{12}{}
  \put(54,15){\makebox(0,0){$\gamma_{L-1}=\alpha_{L}$}}
  \varrow{60}{12}{$\beta_L$}
  \harrow{60}{12}{}
  \put(66,15){\makebox(0,0){$\gamma_L=\alpha_1$}}
  \varrow{60}{0}{$\delta_L$}
\end{picture}
\end{center}
\caption{One layer which corresponds to the transfer matrix.
         The value of arrows incoming to the layer is equal to the
         number of those outgoing from the layer.
         When a bond is shared by two vertices the summation for
         the variable on the bond is taken.
         \label{fig:tmtx}}
\end{figure}

Because there exists the ice rule for the 19-vertex model, the
transfer matrix, by which the partition function $Z$
is expressed, becomes a block diagonal form.
This block diagonal form is obtained in terms of classification
by the value of arrows incoming to one
layer corresponding to the transfer matrix.
The layer with length $L$ is shown in Fig.\ref{fig:tmtx};
$L$ is the width of the system.
We consider the system
under periodic boundary conditions
without frustration in this and next sections.

From the ice rule shown in Eq.(\ref{eq:ice}), we have
\begin{equation}
  \sum_{i=1}^L (\alpha_i-\beta_i-\gamma_i+\delta_i)=0 .
  \label{eq:rowice}
\end{equation}
Because of the following relations
\begin{eqnarray}
  & & \gamma_i = \alpha_{i+1}, ~~ ( 1 \leq i \leq L ) \\
  & & \alpha_{L+1} = \alpha_1 , \label{eq:pbc}
\end{eqnarray}
Eq.(\ref{eq:rowice}) becomes
\begin{equation}
  \sum_{i=1}^L \beta_i = \sum_{i=1}^L \delta_i
  \label{eq:arrowcons}  .
\end{equation}
Here we note that Eq.(\ref{eq:pbc})
comes from the periodic boundary conditions.
The relation given by Eq.(\ref{eq:arrowcons})
means a conservation law of the total value of arrows.
Let us define the total value of arrows $N$ in one layer as follows:
\begin{equation}
  N \equiv \sum_{i=1}^L \beta_i  .
  \label{eq:totalarrow}
\end{equation}
Now the whole transfer matrix of the 19-vertex model is classified
by $N$ and hence has a block diagonal form.
We will explain this in detail in the followings.

By using the property (\ref{eq:arrowcons}),
we reduce an amount of calculation in
the DMRG method.
We apply the infinite system method of the DMRG framework to the
19-vertex model.
In addition to the vertex weight $W(v_i)$ whose values are provided
in Fig.\ref{fig:19vertex}, we define a renormalized weight
$W^{(r)}(v_i^{(r)})$ where $v_i^{(r)}$ means a renormalized vertex
and $r$ is the number of a renormalization.
As an initial value, we set $W^{(0)}(v_i^{(0)})$ to be equal to
$W(v_i)$.
The transfer matrix for the total value of arrows, $N$, is composed as
\begin{eqnarray}
 & & T_N^{(r)}(\eta_1,\beta_2,\eta_3,\beta_4|
               \xi_1,\delta_2,\xi_3,\delta_4)
  \nonumber \\
 &=& \sum_{\alpha_1,\cdots,\alpha_4}
          W^{(r)}(v_1^{(r)}(\alpha_1,\eta_1,\alpha_2,\xi_1))
          W(v_2(\alpha_2,\beta_2,\alpha_3,\delta_2)) \nonumber \\
 &\times& W^{(r)}(v_3^{(r)}(\alpha_3,\eta_3,\alpha_4,\xi_3))
          W(v_4(\alpha_4,\beta_4,\alpha_1,\delta_4))  ,
 \label{eq:maket}
\end{eqnarray}
(see Fig.\ref{fig:Tconstruction})
where the total value of arrows $N$ is obtained by
\begin{eqnarray}
  N &=& N_r(\eta_1)+\beta_2+N_r(\eta_3)+\beta_4 \nonumber \\
    &=& N_r(\xi_1)+\delta_2+N_r(\xi_3)+\delta_4 .
  \label{eq:Aconsv}
\end{eqnarray}
We denote arrows for a vertical bond at a renormalized vertex as
$\xi_i$ or $\eta_i$.
The value of arrows included in the renormalized vertex is
denoted by $N_r(\xi_i)$ for $\xi_i$ or $N_r(\eta_i)$ for $\eta_i$;
$N_r(\xi_i)$ and $N_r(\eta_i)$ are equal to $\delta_i$ and
$\beta_i$ at the initial step of the DMRG procedure.

\begin{figure}
  \begin{center}
  \unitlength 8mm
  \begin{picture}(12,8)
    \put(1,2){\framebox(1,1){$\eta_1$}}
    \put(1,5){\framebox(1,1){$\xi_1$}}
    \put(4,2.5){\circle{1}}
    \put(4,2.5){\makebox(0,0){$\beta_2$}}
    \put(4,5.5){\circle{1}}
    \put(4,5.5){\makebox(0,0){$\delta_2$}}
    \put(6,3.5){\framebox(1,1){$\eta_3$}}
    \put(9,4){\circle{1}}
    \put(9,4){\makebox(0,0){$\beta_4$}}
    \put(5,4.5){\oval(8,5)[tr]}
    \put(5,6){\oval(7,2)[tl]}
    \put(5,3.5){\oval(8,5)[br]}
    \put(5,2){\oval(7,2)[bl]}

    \put(4.38,5.18){\line(4,-3){1.6}}
    \put(4.38,2.82){\line(4,3){1.6}}

    \put(7,4){\line(1,0){1.5}}
    \put(2,5.5){\line(1,0){1.5}}
    \put(2,2.5){\line(1,0){1.5}}

    \put(6.6,6){\makebox(0,0){$\psi_{N,k}^{(r)}
                (\xi_1,\delta_2,\eta_3,\beta_4)$}}
    \put(6.6,2){\makebox(0,0){$\psi_{N,k}^{(r)}
                (\eta_1,\beta_2,\eta_3,\beta_4)$}}
  \end{picture}
  \caption{\label{fig:mkdmtx} Construction of the density matrix by
                  a target state $\vec{\psi}_{N,k}^{(r)}$ of the system.
           $\eta_3$ and $\beta_4$ are shared by two eigenvectors.
           Therefore we obtain a conservation rule
           $N_r(\eta_1)+\beta_2=N_r(\xi_1)+\delta_2$. }
  \end{center}
\end{figure}
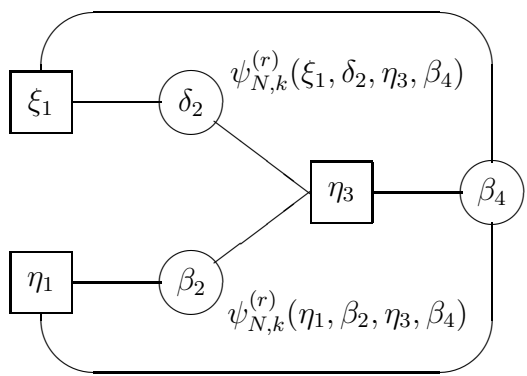

We denote an eigenvector of this transfer matrix by
$\psi_{N,k}^{(r)}(\eta_1,\beta_2,\eta_3,\beta_4)$ which corresponds
to the $k$-th eigenvalue.
As the usual DMRG method, we construct the density matrix
$\hat{\rho}_{N,k}^{(r)}$ as follows:
\begin{equation}
   \rho_{N_r(\eta_1)+\beta_2,k}^{(r)}(\eta_1,\beta_2|\xi_1,\delta_2)
    \equiv \sum_{\eta_3,\beta_4}
     \psi_{N,k}^{(r)}(\eta_1,\beta_2,\eta_3,\beta_4)
     \psi_{N,k}^{(r)}(\xi_1,\delta_2,\eta_3,\beta_4)  .
\end{equation}
Notice that the density matrix is labeled by $N_r(\eta_1)+\beta_2$
not by $N$.
This is due to the fact that $\eta_3$ and $\beta_4$ are shared by
two eigenvectors constructing the density matrix as shown in
Fig.\ref{fig:mkdmtx}.
From Eq.(\ref{eq:Aconsv}), we obtain
$N_r(\eta_1)+\beta_2 = N_r(\xi_1)+\delta_2$,
and therefore the density matrix has a block diagonal form,
as well as the transfer matrix, where each block is classified
by the total value of arrows for the half system.
This property of the density matrix is the reason why we can reduce
a dimension of the transfer matrix by our method introduced in
the present study

In order to construct the renormalized vertex weight $W^{(r+1)}$,
we diagonalize the density matrix
$\hat{\rho}_{N_r(\eta_1)+\beta_2,1}$ and obtain its eigenvectors
$\vec{V}_{N_r(\eta_1)+\beta_2,\eta'_1}$.
In the present study, we use an eigenvector of the transfer matrix
for the largest eigenvalue $k=1$ as a target state.
The renormalized vertex state is labeled by $\eta'_1$ which means
that $\vec{V}_{N_r(\eta_1)+\beta_2,\eta'_1}$ corresponds to the
$\eta'_1$-th eigenvalue of
$\hat{\rho}_{N_r(\eta_1)+\beta_2,1}$.
We determine the upper limit $l$ of $\eta'_1$ as follows:
\begin{eqnarray}
  l \equiv \left\{
     \begin{array}{ll}
        3^{r+2} & (3^{r+2}<m) \\
        m       & (3^{r+2}\geq m)  ,
     \end{array}
  \right.
  \label{eq:rglimit}
\end{eqnarray}
where $m$ is the number of states taken into account for calculation
of the density matrix.
For example, at the initial step of the DMRG method, that is, $r=0$,
the value of $l$ becomes 9 which means the renormalized arrow
$\eta'_1$ includes two arrows.

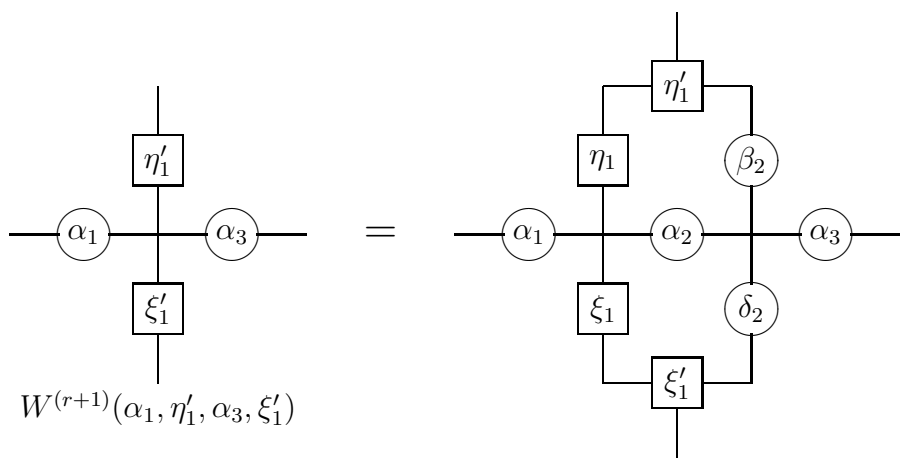
\begin{figure}
\begin{center}
  \setlength{\unitlength}{0.01\textwidth}
  \begin{picture}(72,36)
     \put(12,4){\makebox(0,0){$W^{(r+1)}
     (\alpha_1,\eta'_1,\alpha_3,\xi'_1)$}}

     \harrow{0}{18}{$\alpha_1$}
     \varrowb{12}{18}{$\eta'_1$}
     \harrow{12}{18}{$\alpha_3$}
     \varrowb{12}{6}{$\xi'_1$}

     \put(30,18){\makebox(0,0){\Large $=$}}

     \harrow{36}{18}{$\alpha_1$}
     \varrowb{48}{18}{$\eta_1$}
     \harrow{48}{18}{$\alpha_2$}
     \varrowb{48}{6}{$\xi_1$}

     \varrow{60}{18}{$\beta_2$}
     \harrow{60}{18}{$\alpha_3$}
     \varrow{60}{6}{$\delta_2$}

     \harrowb{48}{30}{$\eta'_1$}
     \put(54,32){\line(0,1){4}}

     \harrowb{48}{6}{$\xi'_1$}
     \put(54,0){\line(0,1){4}}
  \end{picture}
\end{center}
\caption{Reconstruction of $W^{(r)}$ in terms of eigenvectors of
         the density matrix with restriction of the value of arrows.
\label{fig:Wrecon}}
\end{figure}

The last step of the DMRG method for the 19-vertex model is a
construction of the renormalized weight for the vertex as follows:
\begin{eqnarray}
  & & \hspace*{-2em}
   W^{(r+1)}(v_1^{(r+1)}(\alpha_1,\eta'_1,\alpha_3,\xi'_1))
   \nonumber \\
  &=& \sum_{\alpha_2}
      \sum_{\eta_1,\beta_2}\sum_{\xi_1,\delta_2}
      V_{N_r(\eta_1)+\beta_2,\eta'_1}(\eta_1,\beta_2)
                  W^{(r)}(v_1^{(r)}(\alpha_1,\eta_1,\alpha_2,\xi_1))
    \nonumber \\
  &\times&    W(v_2(\alpha_2,\beta_2,\alpha_3,\delta_2))
                  V_{N_r(\xi_1)+\delta_2,\xi'_1}(\xi_1,\delta_2) .
\end{eqnarray}
This last step is illustrated in Fig.\ref{fig:Wrecon}.
The total value of arrows included in the renormalized vertex
becomes
\begin{equation}
   N_{r+1}(\eta'_1)=N_r(\eta_1)+\beta_2  .
\end{equation}
We need also $W^{(r+1)}(v_3^{(r+1)}(\alpha_3,\eta'_3,\alpha_4,\xi'_3))$
when we return to the first step of the DMRG method, but we do not
need to calculate it.
Since in our method for the 19-vertex model the system has a
translational symmetry, we can use $W^{(r+1)}(v^{(r+1)}_1)$
as $W^{(r+1)}(v^{(r+1)}_3)$.
Then we return to the first step represented by Eq.(\ref{eq:maket})
in order to iterate the DMRG procedure.
By iterating this procedure for the 19-vertex model with
the restriction of the total value of arrows as mentioned above,
we are able to make the system size $L$ increase systematically.
The advantage of this method is that the dimension of the transfer
matrix decreases enormously by considering the conservation law of
the value of arrows in this DMRG method.
However we need a large enough value of $m$ in Eq.(\ref{eq:rglimit})
in order to obtain a good accuracy of numerical calculations
for a large system size.
The results for the $m$ dependence of the present method for the BKT
transition are discussed in the following section,
along with the obtained other results.

\section{Results for the conformal anomaly and the smallest
         scaling dimension }
\label{sec:results}
An example of the values of $N_r(\eta_i)$ in the case of $m=35, N=0$ and
$L=12$ is shown in Table \ref{tbl:Nrexample}.
The values of $\eta_i$
in Table \ref{tbl:Nrexample} are put in order
according to the magnitude of eigenvalues of the density matrix.
Because the width of this system, $L$, is twelve, $\eta_i$ represents
five arrows.
Therefore the value of $N_r(\eta_i)$ can take one of
$\{-5,-4, \cdots, 0, \cdots, +4, +5 \}$ as listed in the first column
of Table \ref{tbl:Nrexample}.
Since we set $m=35$, there is no eigenvalue of the
density matrix with $N_r(\eta_i)= \pm 5$.
These are shown by crosses in Table \ref{tbl:Nrexample}.
When the value of arrows is $-4$, we have only one eigenvalue,
which is the 31-st eigenvalue of the density matrix.
This is expressed as $N_r(31)=-4$
in the third line of
Table \ref{tbl:Nrexample}.
In the same way as for $N_r(\eta_i)=-4$, we have two eigenvalues with
$N_r(\eta_i)=-3$; $N_r(12)=-3$ and $N_r(26)=-3$ as shown in
the fourth line
of Table \ref{tbl:Nrexample}.
For other values of $N_r(\eta_i)$, a distribution of $\eta_i$ is
shown in the same manner explained for $N_r(\eta_i)=-5$,$N_r(\eta_i)=-4$
and $N_r(\eta_i)=-3$.
The largest eigenvalue corresponds to $N_r(1)=0$.
For the case of $m=35$ given in Table \ref{tbl:Nrexample},
the smallest eigenvalue corresponds to $ N_r(34)=+1$
and $N_r(35)=-1$.
These two states are degenerate.
In general those two states with $N_r(\eta_i) \neq 0$ are degenerate;
we have an eigenstate with $N_r(\eta_i)$ and an
eigenstate with $-N_r(\eta_i)$ always.
Therefore an eigenstate distribution of the density matrix is
symmetric with respect to $N_r(\eta_i)$.

The dimension of a matrix which has to be diagonalized in the DMRG
method is now enormously reduced by the method mentioned in the
previous section.
Therefore it becomes possible to handle large values of $m$
by using our computer resources.
As an example we show reduction of the dimension of matrices
of which the largest eigenvalues have to be calculated.
In the case of $N=0, m=35$ and $L=12$,
the dimension of a matrix is
$3^{12}=531,441$ for a simple transfer matrix method.
The dimension of the matrix reduces down to
$(35\times3)^2=11,025$ for
the usual DMRG method \cite{White,Nishino}.
The dimension of the matrix is now 1,545 for
our method introduced in the present study.

\begin{figure}
\begin{center}
\epsfile{file=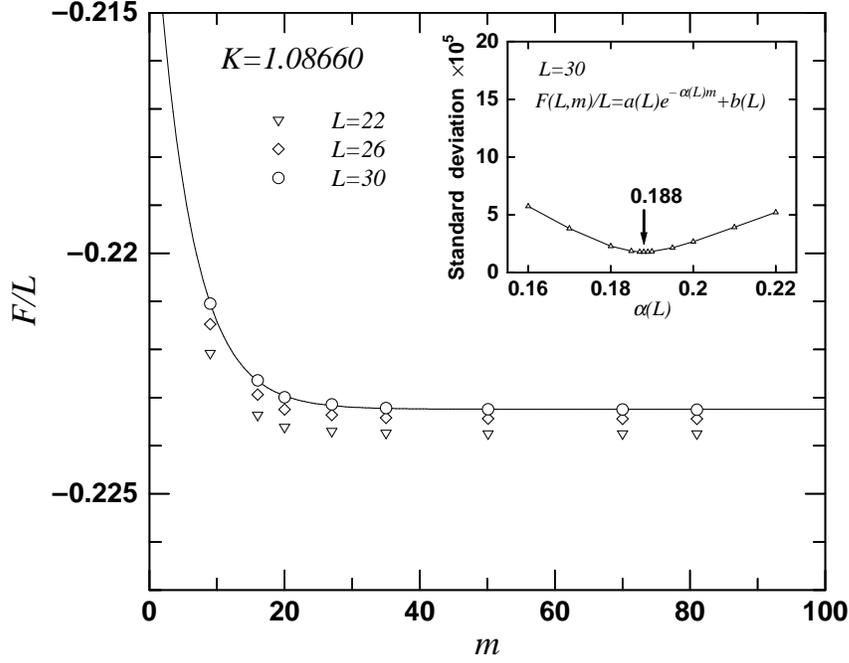,width=0.7\textwidth}
\end{center}
\caption{\label{fig:fmdep}
         $m$ dependence of the free energy obtained by our
         method.
         We use three kinds of fitting functions and obtain the
         smallest standard deviation by an exponential function.
         }
\end{figure}

The largest eigenvalue of the transfer matrix
belongs to the block with $N=0$ in the whole
transfer matrix.
Hence, the largest eigenvalue with $N=0$ is used to calculate
the free energy of the system.
The $m$ dependence of the value of the free energy is shown in
Fig.\ref{fig:fmdep}.
We obtain a good convergence with the value of $m$.
In the case of $L=30$, calculations with $m=3^{14}=4,782,969$ give
an exact diagonalization of the transfer matrix.
Three kinds of fitting functions are used to fit the $m$ dependence.
We obtain the smallest value of standard deviation by using
an exponential function.
These results mean that the largest eigenvalue of the transfer
matrix exponentially converges to an  exact value by increasing $m$
even at the BKT transition point in our method.

\begin{figure}
\begin{center}
\epsfile{file=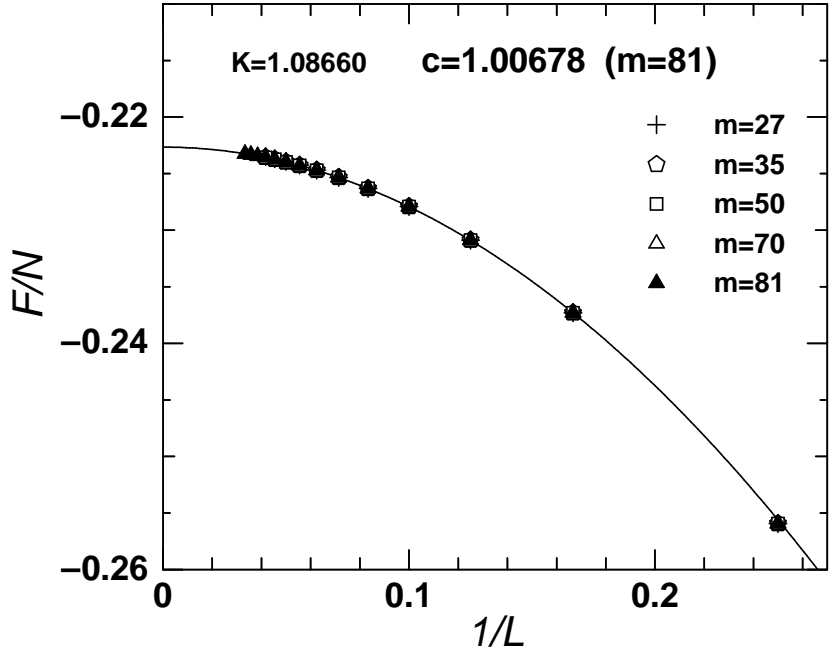,width=0.7\textwidth}
\end{center}
\caption{\label{fig:fldep}
         Size dependence of the free energy of the 19-vertex model
         at $K=1.0866$.
         Estimated value of the conformal anomaly from this result
         is $c=1.006(1)$.
        }
\end{figure}

The size dependence of the free energy $F$ at $K=1.0866$, which is
regarded as the BKT transition point by Knops et al. \cite{Knops},
is shown in Fig.\ref{fig:fldep}.
We use the following size dependence \cite{Blote} of the free energy
for the system with periodic boundary conditions:
\begin{equation}
F/L \sim f_{\infty} - \frac{\pi c}{6 L^2} ,
\label{eq:conformal}
\end{equation}
where $f_{\infty}$ denotes the free energy per site in the
thermodynamic limit.
We obtain the value of the conformal anomaly as $c=1.006(1)$, which is
consistent with the value of $c$ at the critical region.

\begin{figure}
\begin{center}
\epsfile{file=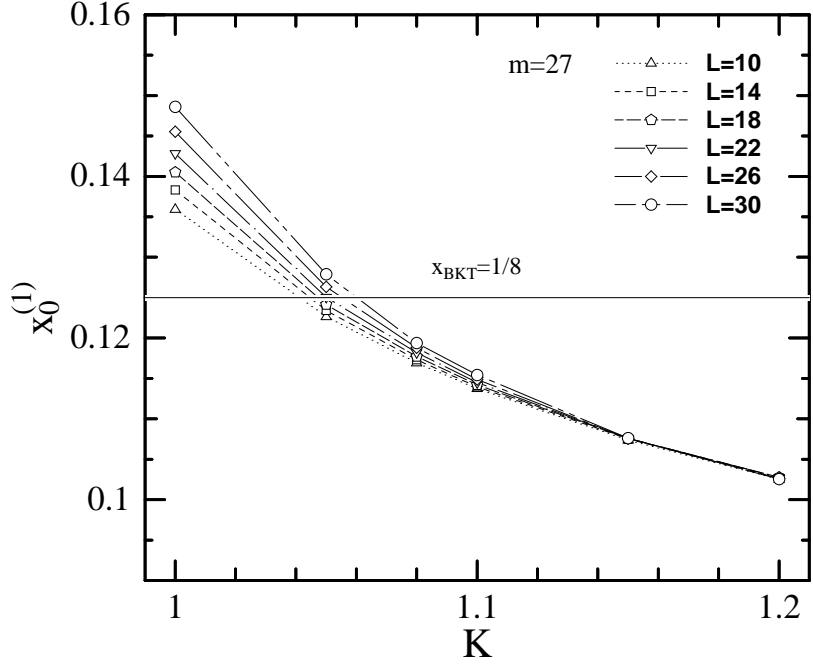,width=0.7\textwidth}
\end{center}
\caption{\label{fig:x1kdep}Interaction dependence of the smallest
         scaling dimension $x_0^{(1)}$}.
\end{figure}

The second largest eigenvalue of the transfer matrix belongs
to a block with $N=1$.
Hence the smallest scaling dimension which is equivalent to $\eta/2$
is estimated by
\begin{equation}
  \eta/2 = x_0^{(1)}
  = \frac{L}{2\pi}\ln \left(
        \frac{\lambda_0^{(0)}}{\lambda_0^{(1)}}
        \right) ,
\label{eq:x1}
\end{equation}
where $\lambda_k^{(n)}$ expresses the $k$-th eigenvalue in the
$N=n$ sector of the transfer matrix \cite{Knops}.
The critical index $\eta$ describes an algebraical decay of a
correlation function in a critical region of interaction.
At the BKT transition point, the value of $\eta$ has to be
$\frac{1}{4}$.
An interaction dependence of $x_0^{(1)}$ is shown in
Fig.\ref{fig:x1kdep}.
These results shown in Fig.\ref{fig:x1kdep}
are obtained by using $m=27$ for each $L$.
In Fig.\ref{fig:x1kdep},
we fail to obtain the value $\frac{1}{8}$ of $x_0^{(1)}$
if we use $K=1.0866$:
the value of $\eta/2$ obviously appears to be smaller than 1/8
at $K=1.0866$.

\begin{figure}
\begin{center}
\epsfile{file=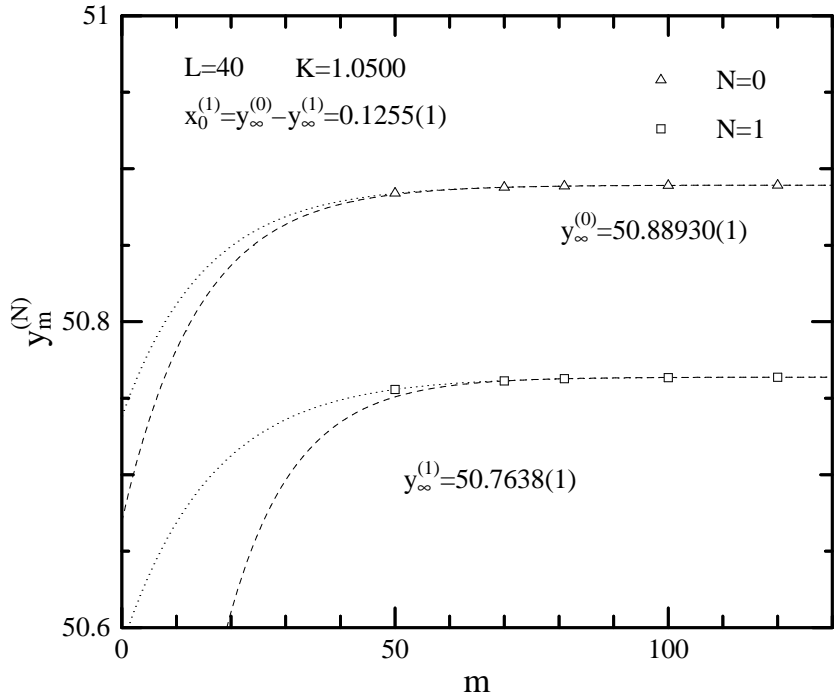,width=0.7\textwidth}
\caption{\label{fig:k1050md}
         $m$ dependences of $y_m^{(N)}$
         at $K=1.0500$ for $L=40$.
         }
\end{center}
\end{figure}

We investigate an $m$-dependence of $x_0^{(1)}$ at $K=1.0866$
and $K=1.0500$ up to $m=120$.
Examples of extrapolations to large values of $m$
by using the exponential function are shown in Fig.\ref{fig:k1050md}.
We define $y_m^{(N)}$ by
\begin{equation}
  y_m^{(N)} \equiv \frac{L}{2\pi} \ln \lambda_{0,m}^{(N)},
  \label{eq:defy}
\end{equation}
where $\lambda_{0,m}^{(N)}$ is the largest eigenvalue of the
transfer matrix with the value of arrows $N$ constructed
by taking account of $m$ states of the density matrix.
To obtain results which are equivalent to an exact diagonalization,
a large value of $m$, i.e., $m=3^{19}$,
for the largest system size $L=40$, which we treat.
Therefore we extrapolate the obtained results $y_m^{(N)}$ to
$m=\infty$ in Fig.\ref{fig:k1050md}.
In Fig.\ref{fig:k1050md}
broken lines are results of fitting by using values obtained from
$m=70, 81, 100$ and 120.
The dotted lines are results of fitting by using values from
$m=50, 70, 81, 100$ and 120.
Difference between results obtained from these two extrapolations
is shown as errors in Fig.\ref{fig:k1050md}.
We evaluate the value of $x_0^{(1)}$ by
$y_{\infty}^{(0)}-y_{\infty}^{(1)}$ and obtain the value,
$x_0^{(1)}=0.1255(1)$ at $K=1.0500$ for $L=40$.

\begin{figure}
\begin{center}
\epsfile{file=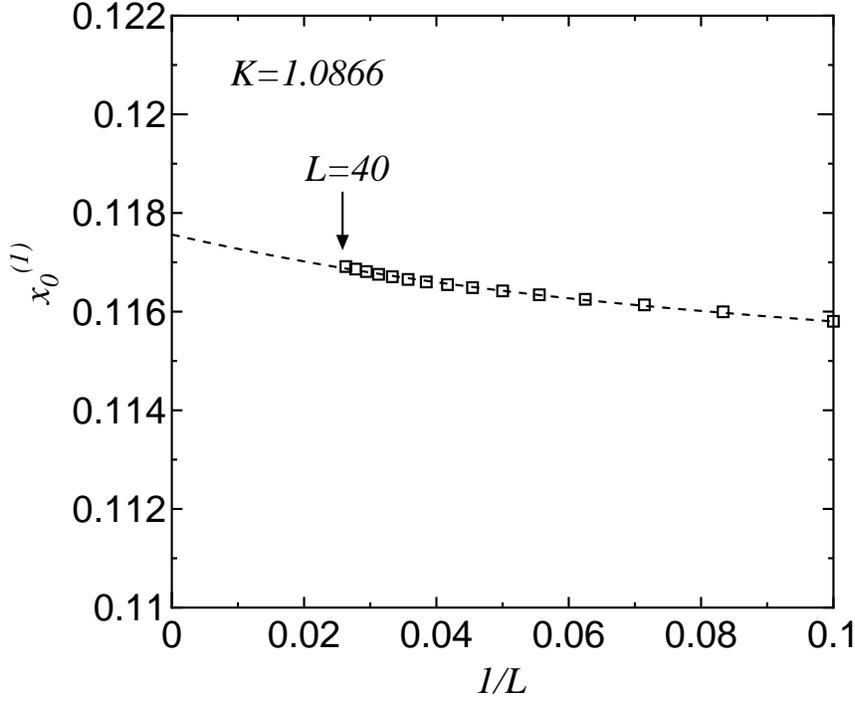,width=0.7\textwidth}
\caption{\label{fig:k10866x1}
         Size dependence of $x_0^{(1)}=\eta/2$
         at $K=1.0866$.
         The value of $x_0^{(1)}$ is expected to be $0.1175(5)$
         for large $L$ which is obviously below 1/8 expected for the
         BKT transition point.
         The broken line shown in this figure is just guide for eyes.}
\end{center}
\end{figure}

\begin{figure}
\begin{center}
\epsfile{file=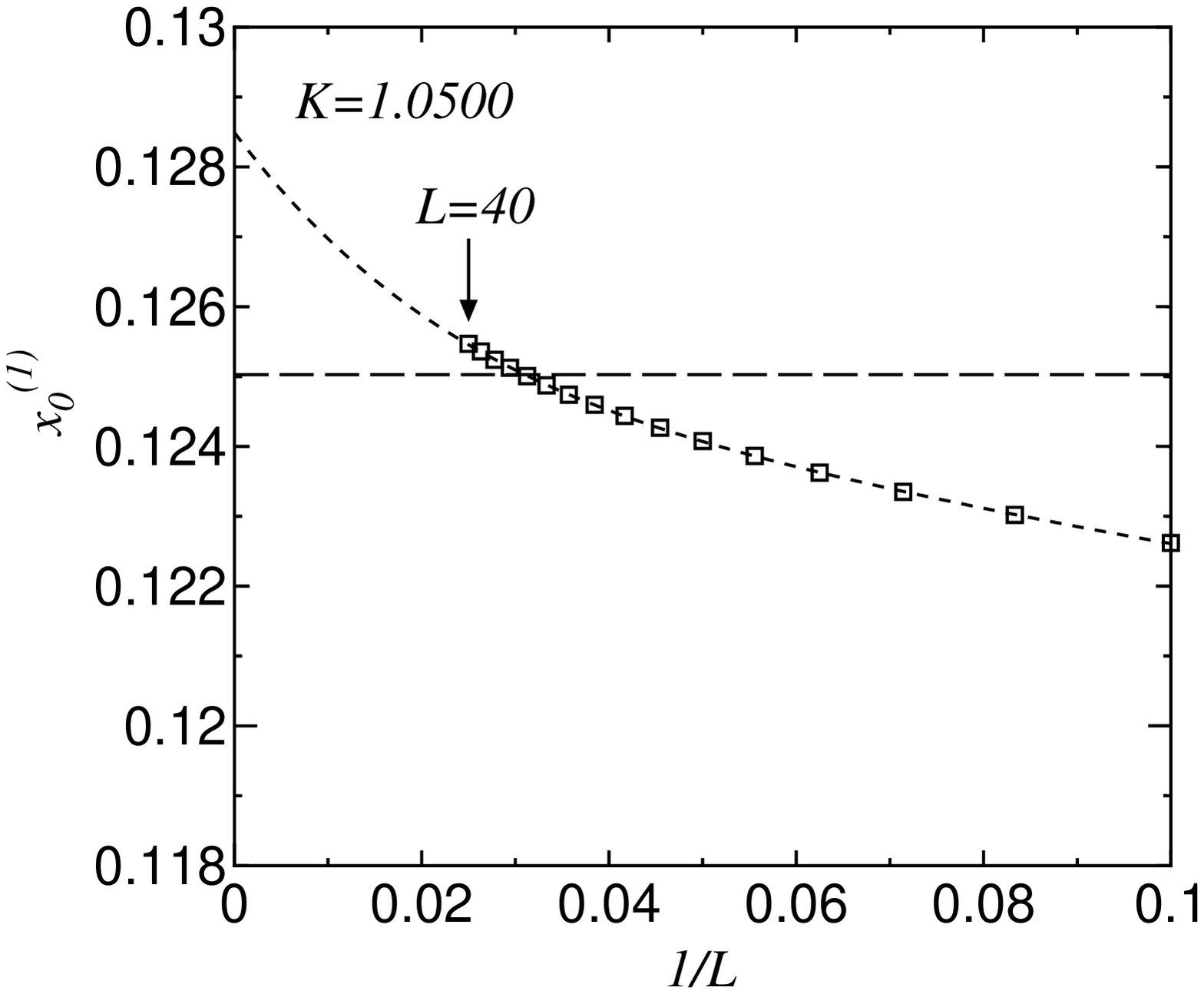,width=0.7\textwidth}
\caption{\label{fig:k10500x1}
         Size dependence of $x_0^{(1)}=\eta/2$ at $K=1.0500$.
         We observe a stronger size dependence than that at $K=1.0866$
         and find that the value is beyond $1/8$.
         The broken line show in this figure is just guide for eyes.
         }
\end{center}
\end{figure}

In Figs.\ref{fig:k10866x1} and \ref{fig:k10500x1},
we show size dependencies of $x_0^{(1)}$ at $K=1.0866$
and $K=1.0500$, respectively.
Every points shown in Figs.\ref{fig:k10866x1}
and \ref{fig:k10500x1}
are values estimated from extrapolations of $m$
as explained above.
We observe that the value of $x_0^{(1)}$ goes to finite value
for large $L$.
The extrapolated value is about $0.1175$ and obviously less than 1/8
at $K=1.0866$.
Therefore $K=1.0866$ is in the critical region.
On the other hand, it is found that the size dependence of
$x_0^{(1)}$ at $K=1.0500$ is stronger than that at $K=1.0866$
as shown in Fig.\ref{fig:k10500x1}.
In addition, the value of $x_0^{(1)}$ increases beyond 1/8
as the system size $L$ increases.
This result indicates that $K=1.0500$ is out of the critical region.

Thus these results at $K=1.0866$ and $K=1.0500$
suggest that the critical value of $K$ of the
19-vertex model is in between $K=1.0500$ and $K=1.0866$.
To determine a precise value of critical point $K_c$ is
an interesting problem.
However this problem is beyond the purpose of the present study.
A further investigation for this point is left as a future problem.

\section{Summary}
\label{sec:summary}
We have succeeded to reduce enormously the dimension of matrix to be
diagonalized in the DMRG method for the 19-vertex model by
using the ice rule.
It has been found that the $m$-dependence of the free energy shows
an exponential convergence even near the BKT transition point by
using our method.
An accurate result is obtained for the conformal anomaly, i.e.,
$c=1.006(1)$
near the BKT transition point by means of the present approach.

From the size dependence of the smallest scaling dimension
$x_0^{(1)}(=\eta/2)$, we have found that $K=1.0866$ belongs to
the critical region.
However, the estimated value of $\eta/2$ is smaller than 1/8.
On the other hand, our results for $K=1.0500$ suggest that
the value of $\eta/2$ is larger than 1/8.
These results means that the critical point of the 19-vertex
model is in between $K=1.0500$ and $K=1.0866$.
A further investigation for this point is left as a future problem.

\acknowledgements
This works partially supported by Grant-In-Aid for Scientific
Research from the Ministry of Education, Science and Culture,
Japan and also by the Computer Center of Tohoku University.

\begin{table}
\begin{center}
 \caption{An example of $N_r(\eta_i)$ structure in the case of
          $m=35, N=0$ and $L=12$. \label{tbl:Nrexample}}
\begin{tabular}{c||ccccccc}
$N_r(\eta_i)$ & & & & $\eta_i$ & & &  \\
\hline
  $-5$ & $\times$ & $\times$ & $\times$ & $\times$ & $\times$
     & $\times$ & $\times$ \\
  $-4$ & 31 &  $\times$ &  $\times$ &  $\times$ &  $\times$
     & $\times$ & $\times$ \\
  $-3$ & 12 & 26 &  $\times$ &  $\times$ &  $\times$ &  $\times$
     & $\times$  \\
  $-2$ & 4  & 11 & 19 & 28 &  $\times$ &  $\times$ & $\times$  \\
  $-1$ & 2  & 8  & 15 & 17 & 23 & 33 & 34 \\
   0 & 1  & 6  & 9  & 14 & 21 & 22 & 29 \\
  $+1$ & 3  & 7  & 16 & 18 & 24 & 32 & 35 \\
  $+2$ & 5  & 10 & 20 & 27 &  $\times$ &  $\times$ &  $\times$ \\
  $+3$ & 13 & 25 &  $\times$ &  $\times$ &  $\times$ &  $\times$
     &  $\times$ \\
  $+4$ & 30 &  $\times$ &  $\times$ &  $\times$ &  $\times$
     & $\times$ &  $\times$ \\
  $+5$ &  $\times$ & $\times$ & $\times$ & $\times$ & $\times$
     & $\times$ & $\times$ \\
\end{tabular}
\end{center}
\end{table}

\begin{figure}
\vspace*{5mm}
\begin{center}
\setlength{\unitlength}{0.007\textwidth}
\begin{picture}(96,160)
  \thicklines
  \footnotesize

  \newcommand{\rarrow}[2]{
     \setcounter{posix}{#1}
     \put(\value{posix},#2){\vector(1,0){6}}
     \addtocounter{posix}{6}
     \put(\value{posix},#2){\line(1,0){2}}
  }
  \newcommand{\larrow}[2]{
     \setcounter{posix}{#1}
     \put(\value{posix},#2){\vector(-1,0){6}}
     \addtocounter{posix}{-6}
     \put(\value{posix},#2){\line(-1,0){2}}
  }
  \newcommand{\uarrow}[2]{
     \setcounter{posiy}{#2}
     \put(#1,\value{posiy}){\vector(0,1){6}}
     \addtocounter{posiy}{6}
     \put(#1,\value{posiy}){\line(0,1){2}}
  }
  \newcommand{\darrow}[2]{
     \setcounter{posiy}{#2}
     \put(#1,\value{posiy}){\vector(0,-1){6}}
     \addtocounter{posiy}{-6}
     \put(#1,\value{posiy}){\line(0,-1){2}}
  }

  \larrow{12}{152}
  \put(12,152){\line(0,1){8}}
  \put(12,152){\line(1,0){8}}
  \uarrow{12}{144}
  \put(12,140){\makebox(0,0){$W(1)=K/2$}}

  \larrow{36}{152}
  \darrow{36}{160}
  \put(36,152){\line(1,0){8}}
  \put(36,152){\line(0,-1){8}}
  \put(36,140){\makebox(0,0){$W(2)=K/2$}}

  \rarrow{52}{152}
  \put(60,152){\line(0,1){8}}
  \put(60,152){\line(1,0){8}}
  \darrow{60}{152}
  \put(60,140){\makebox(0,0){$W(3)=K/2$}}

  \rarrow{76}{152}
  \uarrow{84}{152}
  \put(84,152){\line(1,0){8}}
  \put(84,152){\line(0,-1){8}}
  \put(84,140){\makebox(0,0){$W(4)=K/2$}}

  \put(12,120){\line(-1,0){8}}
  \put(12,120){\line(0,1){8}}
  \rarrow{12}{120}
  \uarrow{12}{112}
  \put(12,108){\makebox(0,0){$W(5)=K/2$}}

  \put(36,120){\line(-1,0){8}}
  \darrow{36}{128}
  \rarrow{36}{120}
  \put(36,120){\line(0,-1){8}}
  \put(36,108){\makebox(0,0){$W(6)=K/2$}}

  \put(60,120){\line(-1,0){8}}
  \put(60,120){\line(0,1){8}}
  \larrow{68}{120}
  \darrow{60}{120}
  \put(60,108){\makebox(0,0){$W(7)=K/2$}}

  \put(84,120){\line(-1,0){8}}
  \uarrow{84}{120}
  \larrow{92}{120}
  \put(84,120){\line(0,-1){8}}
  \put(84,108){\makebox(0,0){$W(8)=K/2$}}

  \rarrow{4}{88}
  \put(12,88){\line(0,1){8}}
  \rarrow{12}{88}
  \put(12,88){\line(0,-1){8}}
  \put(12,76){\makebox(0,0){$W(9)=K/2$}}

  \larrow{36}{88}
  \put(36,88){\line(0,1){8}}
  \larrow{44}{88}
  \put(36,88){\line(0,-1){8}}
  \put(36,76){\makebox(0,0){$W(10)=K/2$}}

  \put(60,88){\line(-1,0){8}}
  \uarrow{60}{88}
  \put(60,88){\line(1,0){8}}
  \uarrow{60}{80}
  \put(60,76){\makebox(0,0){$W(11)=K/2$}}

  \put(84,88){\line(-1,0){8}}
  \darrow{84}{96}
  \put(84,88){\line(1,0){8}}
  \darrow{84}{88}
  \put(84,76){\makebox(0,0){$W(12)=K/2$}}

  \rarrow{4}{56}
  \uarrow{12}{56}
  \rarrow{12}{56}
  \uarrow{12}{48}
  \put(12,44){\makebox(0,0){$W(13)=K^2/4$}}

  \rarrow{28}{56}
  \darrow{36}{64}
  \rarrow{36}{56}
  \darrow{36}{56}
  \put(36,44){\makebox(0,0){$W(14)=K^2/4$}}

  \larrow{60}{56}
  \uarrow{60}{56}
  \larrow{68}{56}
  \uarrow{60}{48}
  \put(60,44){\makebox(0,0){$W(15)=K^2/4$}}

  \larrow{84}{56}
  \darrow{84}{64}
  \larrow{92}{56}
  \darrow{84}{56}
  \put(84,44){\makebox(0,0){$W(16)=K^2/4$}}

  \larrow{12}{24}
  \darrow{12}{32}
  \rarrow{12}{24}
  \uarrow{12}{16}
  \put(12,12){\makebox(0,0){$W(17)=K^2/4$}}

  \rarrow{28}{24}
  \uarrow{36}{24}
  \larrow{44}{24}
  \darrow{36}{24}
  \put(36,12){\makebox(0,0){$W(18)=K^2/4$}}

  \put(60,24){\line(-1,0){8}}
  \put(60,24){\line(0,1){8}}
  \put(60,24){\line(1,0){8}}
  \put(60,24){\line(0,-1){8}}
  \put(60,12){\makebox(0,0){$W(19)=1$}}

\end{picture}
\end{center}
\caption{19 vertices permitted by the ice rule generalized to include
         no arrow on a bond
 \label{fig:19vertex}}
\end{figure}
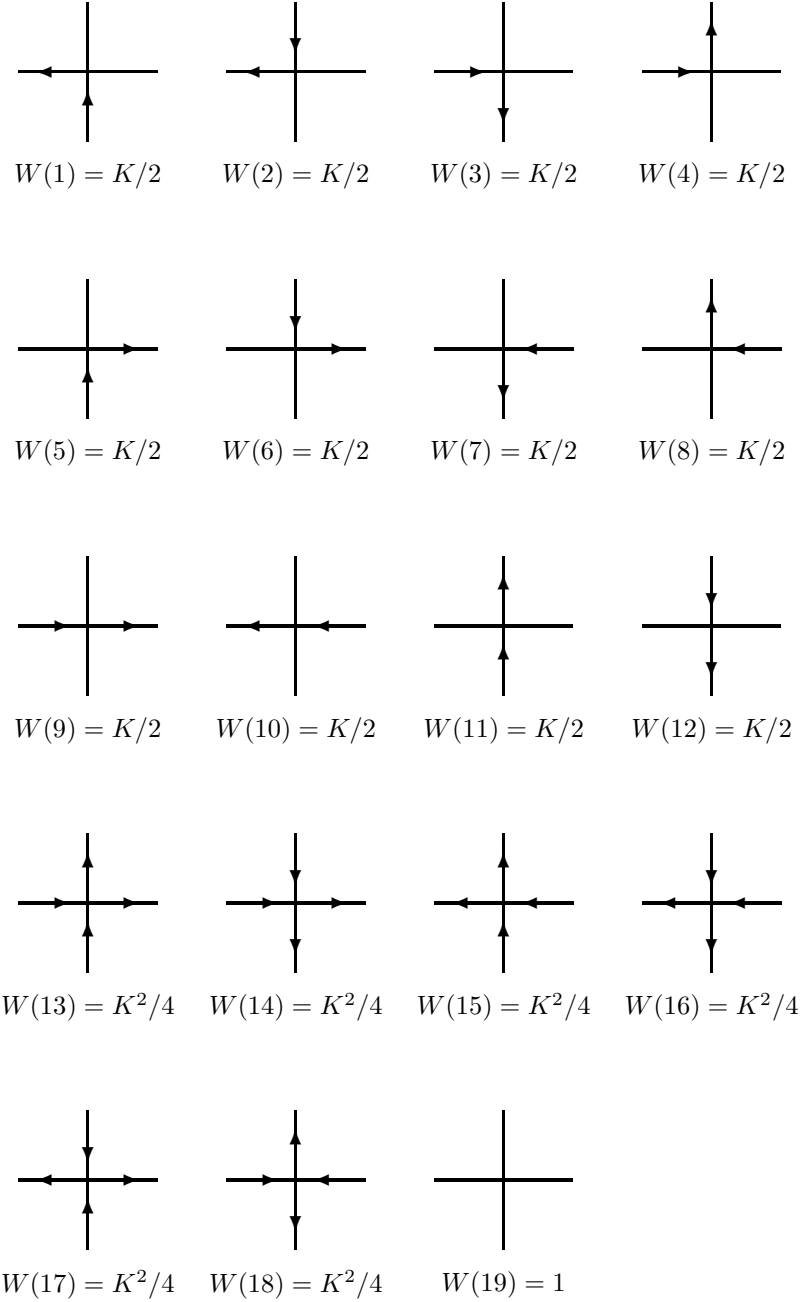

\begin{figure}
\begin{center}
\setlength{\unitlength}{0.011\textwidth}
\begin{picture}(60,24)
  \footnotesize
  \harrow{0}{12}{$\alpha_1$}
  \varrowb{12}{12}{$\eta_1$}
  \harrow{12}{12}{$\alpha_2$}
  \varrowb{12}{0}{$\xi_1$}

  \varrow{24}{12}{$\beta_2$}
  \harrow{24}{12}{$\alpha_3$}
  \varrow{24}{0}{$\delta_2$}

  \varrowb{36}{12}{$\eta_3$}
  \harrow{36}{12}{$\alpha_4$}
  \varrowb{36}{0}{$\xi_3$}

  \varrow{48}{12}{$\beta_4$}
  \harrow{48}{12}{$\alpha_1$}
  \varrow{48}{0}{$\delta_4$}

\end{picture}
\end{center}
\caption{Transfer matrix composed by weights $W$ and $W^{(r)}$
\label{fig:Tconstruction}}
\end{figure}
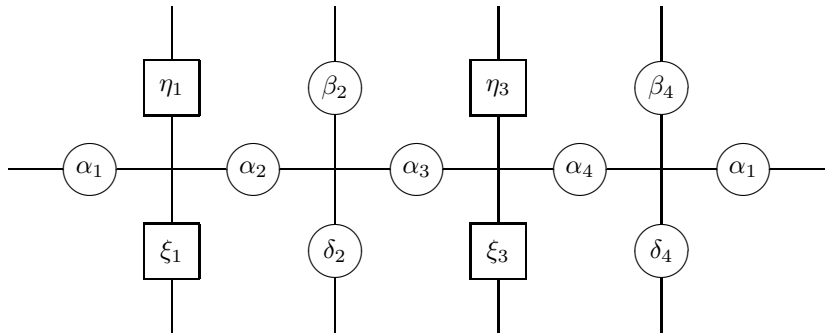

\end{document}